\documentclass[twocolumn,pra,aps,showpacs,superscriptaddress]{revtex4-1}
\usepackage[english]{babel}
\usepackage[utf8]{inputenc}
\usepackage{url,psfrag,graphicx}
\usepackage{dcolumn}
\usepackage{amsmath,amssymb,amsthm}
\usepackage{bm}
\usepackage{pstricks}
\usepackage{hyperref}
\usepackage{epsfig,placeins}

\def\ket|#1>{|#1\rangle}
\def\bra<#1|{\langle #1|}
\def\braket<#1|#2|#3>{\langle #1 | #2 | #3 \rangle}
\def\<{\langle}
\def\>{\rangle}
\def\{{\lbrace}
\def\}{\rbrace}
\def\({\left(}
\def\){\right)}
\def\beq{\begin{equation}}
\def\eeq{\end{equation}}

\def\nn{\nonumber}

\def\H{\mathcal{H}}
\def\Tr{\text{Tr}}
\def\ve{\varepsilon}

\def\tE{{\widetilde E}}
\def\Ez{E_{A,0}}

\begin{document}

\title{Ergotropy and entanglement in critical spin chains}

\author{Begoña Mula}
\affiliation{Departamento de Física Fundamental, Universidad Nacional
  de Educación a Distancia (UNED), Madrid, Spain}
\affiliation{Departamento de Matemáticas, Universidad Carlos III de
  Madrid, Leganés, Spain}

\author{Eva M. Fernández}
\affiliation{Departamento de Física Fundamental, Universidad Nacional
  de Educación a Distancia (UNED), Madrid, Spain}

\author{José E. Alvarellos}
\affiliation{Departamento de Física Fundamental, Universidad Nacional
  de Educación a Distancia (UNED), Madrid, Spain}

\author{Julio J. Fernández}
\affiliation{Departamento de Física Fundamental, Universidad Nacional
  de Educación a Distancia (UNED), Madrid, Spain}

\author{David García-Aldea}
\affiliation{Departamento de Física Fundamental, Universidad Nacional
  de Educación a Distancia (UNED), Madrid, Spain}

\author{Silvia N. Santalla}
\affiliation{Departamento de Física \&\ GISC, Universidad Carlos III
  de Madrid, Leganés, Spain}

\author{Javier Rodríguez-Laguna}
\affiliation{Departamento de Física Fundamental, Universidad Nacional
  de Educación a Distancia (UNED), Madrid, Spain}

\begin{abstract}
  A subsystem of an entangled ground state is in a mixed state. Thus,
  if we isolate this subsystem from its surroundings we may be able to
  extract work applying unitary transformations, up to a maximal
  amount which is called ergotropy. Once this work has been extracted,
  the subsystem will still contain some {\em bound energy} above its
  local ground state, which can provide valuable information about the
  entanglement structure. We show that the bound energy for half a
  free fermionic chain decays as the square of the entanglement
  entropy divided by the chain length, thus approaching zero for large
  system sizes, and we conjecture that this relation holds for all 1D
  critical states.
\end{abstract}

\date{January 15, 2023}

\maketitle

\section{Introduction}

Quantum thermodynamics applies the core concepts of quantum
information theory \cite{Gemmer.04,Binder.18,Bera.19} to design
optimal nanoscale devices, such as quantum thermal machines
\cite{Scully.03, Ghosh.19,Fernandez.22}. A very fruitful concept is
that of {\em ergotropy} \cite{Allahverdyan.04,Alicki.13}, i.e. the
maximal work that can be reversibly extracted from a mixed state,
which is a crucial tool in order to build efficient quantum batteries
\cite{Campaioli.17,Andolina.19,Rosa.20}. Indeed, ergotropy is known to
be strongly influenced by the presence of quantum correlations of
different types \cite{Francica.17,Francica.20,Tirone.21,
  Touil.22,Francica.22}. Of course, if we lift the reversibility
constraint, we may use quantum measurements to extract work in an
optimal way \cite{Manzano.18,Solfanelli.19}.

Yet, the connection works in both directions, and we may employ
quantum thermodynamics to characterize the entanglement structure of
a quantum system. As it is well known, a subsystem of a ground state (GS)
is usually not in its local ground state. Instead, it must be
described by a reduced density matrix, which can be expressed as a
thermal density matrix under a certain entanglement Hamiltonian (EH),
which need not coincide with the local one
\cite{Bisognano.75,Bisognano.76}. Notice that the EH allows us to
describe the entanglement structure of complex quantum states in
thermal terms. Both the EH and its eigenvalues, which define the
entanglement spectrum (ES) \cite{Li.08}, have provided invaluable
insight to characterize the entanglement structure of the low energy
states of quantum manybody systems
\cite{Peschel.99,Chung.00,Chung.01,Peschel.03,Casini.09,Casini.11,
  Peschel.17,Santos.18,Samos.20}, in some cases exploiting their
conformal invariance \cite{Cardy.16,Tonni.18,Samos.21}.

In this work we introduce the notion of {\em subsystem ergotropy}
within a GS in order to characterize its entanglement structure
through the analysis of the energetic relations between a subsystem
$A$ and its environment $B$. The expected value of the local energy of
any subsystem will typically exceed its own GS energy, and the
subsystem ergotropy is defined as the part that can be extracted in
the form of work. Our analysis will focus on a few simple quantum
many-body systems, starting with a detailed analysis of free fermionic
chains, and extending our study to other critical spin chains. In all
the considered cases, we benefit from the constraints imposed by
conformal invariance on the reduced density matrix. We show that, once
the maximal work has been extracted, the remaining bound energy
presents universal scaling as the square of the entanglement entropy
of the block divided by the system size, thus approaching zero for
large system sizes.

This article is organized as follows. Section \ref{sec:theory}
develops the basic theoretical background, combining tools from
quantum thermodynamics and quantum information theory. Then we show
our analytical and numerical calculations for a free fermionic chain
in Sec. \ref{sec:ff}. Other critical spin chains, such as the Ising
model in a transverse field or the Heisenberg model, are briefly
considered in \ref{sec:interacting}. The article ends with a section
describing our conclusions and suggestions for further work.


\section{Theoretical background}
\label{sec:theory}

\subsection{Ergotropy of generic mixed states}

The {\em ergotropy} $W$ of a mixed state $\rho$ with respect to a
Hamiltonian $H$ can be defined as the maximal amount of work that can
be extracted from the state by applying unitary operations
\cite{Allahverdyan.04,Alicki.13}, i.e.

\beq
W\equiv \max_U \(\Tr(\rho\, H)-\Tr(U\rho U^\dagger\, H)\),
\label{eq:ergotropy}
\eeq
where $U$ is any unitary transformation. Alternatively, it can be
shown \cite{Allahverdyan.04} that the ergotropy corresponds to the
maximal work that can be reversibly extracted from the system, but the
former characterization suits our purposes better. A state defined by
a density matrix $\rho$ is called {\em passive} with respect to $H$
when its ergotropy is zero, i.e. when we can not extract any work from
it by performing unitary operations. In that case, the eigenstates of
$H$ and $\rho$ must be {\em aligned} such that the highest probability
state of $\rho$ will correspond to the lowest eigenstate of $H$, and
so on. Thermal states built on $H$, written as $\rho=Z^{-1}\exp(-\beta
H)$ with $\beta=1/T$ the inverse temperature (we assume $k_B=1$) and
$Z$ the normalization factor, are always passive with respect to their
Hamiltonian, but the converse is not true. In other words: all thermal
states are passive, but not all passive states are thermal.

Let $p_0\geq p_1 \geq \cdots \geq p_{N-1}$ be the eigenvalues of
$\rho$, where $N$ is the dimension of the Hilbert space. Similarly,
let $E_0\leq E_1 \leq \cdots \leq E_{N-1}$ denote the eigenvalues of
$H$ in ascending order, and let $E=\Tr(\rho H)$ be the expected value
of the energy of the system. Now, let us define the passivized state,

\beq
\tilde\rho\equiv U\rho U^\dagger,
\label{eq:tilderho}
\eeq
with $U$ the unitary operator implicitly defined in
Eq. \eqref{eq:ergotropy}. Naturally, the spectra of both density
matrices must coincide,

\beq
\text{Sp}(\rho)=\text{Sp}(\tilde\rho)=\{p_k\}_{k=0}^{N-1},
\eeq
Since the passive energy $\widetilde E\equiv \Tr(\tilde\rho H)$ must be
minimal among all density matrices with the same spectrum, we deduce
that the maximal probability, $p_0$, must share eigenstate with the
the GS energy of $H$, $E_0$; the second probability, $p_1$, with the
first excited state, $E_1$, and so on. Therefore,

\beq
\widetilde E = \sum_{k=0}^{N-1} p_k E_k,
\eeq
and degeneracies do not pose any complications. The ergotropy is given
by

\beq
W\equiv E-\widetilde E \leq E-E_0.
\eeq
Notice that since we have chosen a common basis of eigenvectors of $H$
and $\tilde\rho$, the two operators must commute,
$[H,\tilde\rho]=0$. In general, this density matrix $\tilde\rho$ need
not be {\em thermal} for $H$, i.e. it may not be written as
$\tilde\rho \approx Z^{-1}\exp(-\beta H)$ for any value of $\beta$.

\subsection{Subsystem ergotropy}

Let us consider a quantum system on a composite Hilbert space
$\H=\H_A\otimes\H_B$, with Hamiltonian $H$,

\beq
H=H_A\otimes I_B+I_A\otimes H_B+H_{AB} \equiv H_0 + H_{AB},
\label{eq:ham}
\eeq
where $H_{\{A,B\}}$ acts on $\H_{\{A,B\}}$ respectively, and $H_{AB}$
will be called the {\em interaction Hamiltonian}. Of course, this
decomposition is not unique, and we will assume that $H_{AB}$ has been
chosen {\em as small as possible} in some norm. Let $\ket|\Psi>$ be
the (non-degenerate) GS energy of $H$, which can always be written as
a Schmidt decomposition

\beq
\ket|\Psi>=\sum_{k=1}^\chi p_k^{1/2} \ket|\phi^A_k>\otimes\ket|\phi^B_k>,
\label{eq:schmidt}
\eeq
where $\ket|\phi^A_k>\in \H_A$, $\ket|\phi^B_k>\in \H_B$ are two
orthornormal sets, $p_k\geq 0$ (also in non-increasing order) and
$\chi\leq \min(\dim(\H_A),\dim(\H_B))$ is the Schmidt number. The
reduced density matrix for part $A$ can be written as

\beq
\rho_A=\sum_{k=1}^\chi p_k \ket|\phi^A_k>\bra<\phi^A_k|.
\label{eq:rdmA}
\eeq
Being positive definite, this matrix can always be written as a
thermal density matrix,

\beq
\rho_A=\exp(-K_A),
\label{eq:EH}
\eeq
where $K_A$ is called the {\em entanglement Hamiltonian} (EH)
associated to part $A$. Of course, $K_A$ need not be equal to $H_A$,
the local Hamiltonian, and this difference will be crucial in what
follows. Also, let us introduce the entanglement spectrum (ES) as the
spectrum of the EH \cite{Li.08}.

\begin{figure}
  \includegraphics[width=8cm]{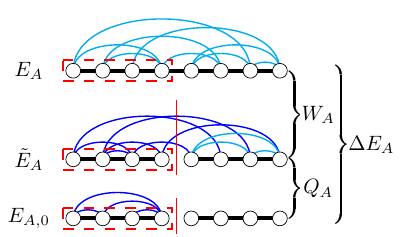}
  \caption{Illustrating the energies involved in our discussion of the
    subsystem ergotropy and their differences. Indeed, $E_A$ denotes
    the expected value of the $H_A$ within the global GS of $H$,
    $\tE_A$ is the minimal energy achieved through unitary operations
    on ${\cal H}_A$ and $E_{A,0}$ is the GS of $H_A$. Moreover,
    $\Delta E_A=E_A-E_{A,0}$ is the {\em excess energy},
    $W_A=E_A-\tE_A$ is the subsystem ergotropy and $Q_A=\tE_A-E_{A,0}$
    is the subsystem bound energy. The blue archs denote entanglement,
    as it is explained in the text. Notice that, in order to define
    these energies, block $A$ must be physically separated from its
    environment.}
  \label{fig:illust}
\end{figure}

Now let us physically separate subsystem $A$ from its environment,
i.e. subsystem $B$, by suddenly quenching $H_{AB}$ to zero. The
subsequent behavior of our subsystem will be described by $H_A$, with
spectrum $\{E_{A,k}\}$, which we may assume to be non-degenerate. We
define the three energies involved in our problem:

\begin{itemize}
  \item $E_A=\<\Psi|H_A\otimes I_B|\Psi\>$, the expected value of
    $H_A$ in the global GS.
  \item $\widetilde E_A=\sum_k p_k E_{A,k}$, the passive energy of the
    system, obtained through unitary transformations.
  \item $E_{A,0}$, the GS of $H_A$.
\end{itemize}

These three energies must be in descending order, $E_A \geq \tE_A \geq
E_{A,0}$. We define the {\em excess energy} as $\Delta E_A\equiv
E_A-E_{A,0}$. The subsystem ergotropy can be computed as

\beq
W_A=E_A-\tE_A,
\eeq
while

\beq
Q_A\equiv \tE_A-\Ez,
\eeq
denotes the amount of energy which is unavailable, which we will call
the {\em subsystem bound energy} \cite{Bera.19}. See Fig.
\ref{fig:illust} for an illustration. The top panel represents the GS
of $H$, and $E_A$ is the energy associated to block $A$. The light
blue archs represent the entanglement links \cite{Singha.20,Singha.21}
which characterize the entanglement structure. We reach the middle
panel applying a suitable unitary operator on block $A$, maximally
reducing its energy to $\tE_A$ while preserving the entanglement
spectrum and, {\em a fortiori}, the amount of entanglement with the
rest of the system, which in this figure is represented by the number
of links leaving $A$. The newly established links are now denoted in
dark blue. Finally, the lowest panel denotes the GS of $H_A$, which is
now disentangled from the environment, with energy $E_{A,0}$.

\subsection{Ergotropy and time evolution}

Once we have split the subsystem $A$ from its environment, it will
evolve under the action of its local Hamiltonian, $H_A$, following von
Neumann's equation,

\beq
i\hbar\; \partial_t \rho_A = [H_A,\rho_A].
\label{eq:von_neumann}
\eeq
Remarkably, this time evolution preserves both the expected value of
the energy, $E_A$, and the full spectrum of the density matrix, even
though the subsequent dynamics can be complex
\cite{Zamora.14,Shimaji.19,Rolph.22}. It is relevant to ask how much
work we can obtain from this time-evolved density matrix employing
unitary transformations, i.e. how the ergotropy evolves after the
split quench. The answer is that the ergotropy is exactly preserved
along the time evolution. A proof of this fact is straightforward. The
time-evolved density matrix for the subsystem after the split can be
written as $\rho_A(t)=V(t)\rho_A(0)V^\dagger(t)$ for some unitary
transformation $V(t)$. The ergotropy of this matrix, defined in
Eq. \eqref{eq:ergotropy}, is exactly the same, because the associated
passivized state, given in Eq. \eqref{eq:tilderho}, is exactly the
same, if we just use the identity

\beq
\tilde\rho_A = U\rho_A(0) U^\dagger = U V^\dagger(t) \rho_A(t) V(t)
U^\dagger,
\eeq
allowing us to define a new unitary transformation, $\tilde
U=UV^\dagger(t)$, such that $\tilde\rho_A=\tilde U \rho_A \tilde
U^\dagger$.  This result implies that the work extraction procedure
need not start immediately after the disconnection between the
subsystem and its environment, as long as the subsequent evolution is
unitary.

\subsection{Interaction energy inequality}

Thus, we can extract work from a subsystem of a composite quantum
state in its GS. Yet, this work should always be less than the
corresponding increase in the energy of the system induced by our
interaction, because otherwise the current system energy would be
lower than the GS energy, $E$. We can prove this result easily. After
the unitary transformation on subsystem $A$ the global system will be
$\ket|\tilde\Psi>$, such that

\beq
\<\tilde\Psi|H|\tilde\Psi\>=\widetilde E = \widetilde E_A + \widetilde E_B +
\widetilde E_{AB},
\eeq
where each term on the rhs corresponds to the expectation value of one
of the three operators, $H_A$, $H_B$ and $H_{AB}$ on
$\ket|\tilde\Psi>$, and we notice that $\widetilde E_B=E_B$. This energy
$\widetilde E \geq E$, the GS energy, which can be decomposed equally,
$E=E_A+E_B+E_{AB}$. Taking into account that $E_A-\widetilde E_A=W_A$, we
obtain

\beq
\widetilde E_{AB}-E_{AB} \geq W_A \geq 0.
\eeq
which implies that the gain through ergotropy must be less or equal
than the loss in the interaction term.


\section{Ergotropy of a free fermionic chain}
\label{sec:ff}

We now particularize the previous calculation to the case of a free
fermionic chain, before extending our results to other critical spin
chains. As we will show, the ergotropy and bound energy of free
fermionic chains can be explicitly computed and present universal
features associated to conformal invariance, in similarity to the
Casimir energy \cite{Cardy.84,Cardy.86,Mula.21}. For simplicity, we
will restrict ourselves to the case in which the block $A$ corresponds
to the left half of the chain.

\subsection{Free fermionic chains}

Let us consider a fermionic chain of $N$ (even) sites with open
boundaries, described by the Hamiltonian

\beq
H_N =-\sum_{i,j=1}^N J_{ij}\; c^\dagger_i c_j,
\label{eq:ffham}
\eeq
where $c^\dagger_i$ and $c_i$ denote the fermionic creation and
annihilation operators on site $i$ and $J_{ij}=\bar J_{ji}$ denotes
the {\em hopping matrix}. We will focus on the homogeneous chain with
open boundaries, whose hopping amplitudes are given by
$J_{ij}=\delta_{i,j\pm 1}$. In this case, the low energy behavior of
the chain can be accurately represented by a conformal field theory
(CFT) \cite{DiFrancesco,Mussardo}.

The GS of Hamiltonian \eqref{eq:ffham} can be obtained through the
eigenvalues $\{\ve_k\}$ (in increasing order) and eigenmodes
$\{U_{k,i}\}$ of the hopping matrix $J_{ij}$, which are usually called
single-body energies and modes, respectively. The spectrum presents
particle-hole symmetry, $\ve_k=-\ve_{N+1-k}$, and the GS is obtained
by filling up the $N/2$ negative energy modes, such that

\beq
E=\sum_{k=1}^{N/2} \ve_k,
\eeq
while the corresponding eigenstate is a Slater determinant determined
by its correlator matrix, defined as

\beq
C_{ij}\equiv\<c^\dagger_ic_j\>=\sum_{k=1}^{N/2} \bar U_{k,i} U_{k,j}.
\label{eq:corr}
\eeq
All the entanglement properties can be determined from matrix
$C$. Indeed, the reduced density matrix of any block $A$ of size
$\ell$ can be obtained diagonalizing the corresponding
$\ell\times\ell$ submatrix, $C_A$. The set $\{\nu^A_k\}$ of
eigenvalues of $C_A$, where each $\nu^A_k \in [0,1]$ determines
uniquely the full ES, will be called {\em entanglement
  occupations}. The von Neumann entropy of block $A$ can be expressed
as \cite{Peschel.03}

\beq
S_A=-\sum_{k=1}^\ell \(\nu^A_k\log(\nu^A_k) +
(1-\nu^A_k)\log(1-\nu^A_k)\).
\eeq
Conformal symmetry fixes the universal part of the entanglement
entropy of a lateral block $A=\{1,\cdots,\ell\}$ of a critical chain
with $N$ sites, \cite{Holzhey.94,Vidal.03,Calabrese.09}

\beq
S_A \approx {c\over 6} \log\({N\over \pi}\sin\({\pi\ell\over N}\)\) +
c',
\label{eq:Scft}
\eeq
where $c=1$ is the central charge of the associated CFT
\cite{DiFrancesco,Mussardo} and $c'$ is a non-universal constant.
Moreover, the EH of a free fermionic chain must also present a
free fermionic form, Eq. \eqref{eq:ffham}, with a different hopping
matrix \cite{Bisognano.75,Bisognano.76},

\beq
\rho_A={1\over Z} \exp(-K_A) =
    {1\over Z} \exp\(-\sum_{i,j=1}^\ell K^A_{ij} c^\dagger_i c_j\).
\eeq
The single-body energies of the EH, ${\cal E}^A_k$, can be obtained from
the entanglement occupations through the Fermi-Dirac expression,

\beq
\nu^A_k = {1\over 1+\exp({\cal E}^A_k)},
\eeq
and they are (approximately) equally spaced, with a level separation
given by the so-called {\em entanglement gap}, ${\cal E}_A \approx
{\cal E}^A_{k+1}-{\cal E}^A_k$, which is known to behave like
\cite{Tonni.18}

\beq
    {\cal E}_A \approx {2\pi^2 \over \log(\gamma N)},
    \label{eq:Sgap}
\eeq
where $\log\gamma\approx 2.3$ is a non-universal constant
\cite{Tonni.18}. Moreover, an approximate inverse relation has been
proposed between the entanglement gap and the entanglement entropy,

\beq
{\cal E}_A S_A \approx {\pi^2\over 3}.
\eeq

\subsection{Casimir energy and free fermions}

Our next aim is to compute the three energies involved in our
calculations: $E_A$, $\tE_A$ and $\Ez$. Let us start with $\Ez$ for
convenience. We proceed to build $H_A$, the hopping matrix for the
block $A$, and obtain its eigenvalues, $\{\ve^A_k\}_{k=1}^{N/2}$, in
increasing order. The GS energy of A is given by

\beq
\Ez = \sum_{k=1}^{N/4} \ve^A_k.
\eeq

An approximate expression for $\Ez$ as a function of
$N$ can be provided \cite{Cardy.84,Cardy.86,Mula.21}

\beq
E_0(N) = -c_0(N-1)-c_B - {c\pi v_F\over 24 N} + O(N^{-2}),
\label{eq:casimir}
\eeq
where we distinguish three terms. The first one, $-c_0(N-1)$, with
$c_0=2/\pi$, is the {\em bulk energy}. The second term,
$-c_B=-(4/\pi-1)$, is the {\em boundary term}. The third one provides
the finite-size correction and is fixed by conformal
invariance. Indeed, $c=1$ is the central charge associated to our
theory and $v_F=2$ is the Fermi velocity. Thus, we have

\beq
\Ez \approx -c_0\({N\over 2}-1\) -c_B - {\pi \over 6N}.
\label{eq:EA0}
\eeq

\begin{figure}
  \includegraphics[width=8cm]{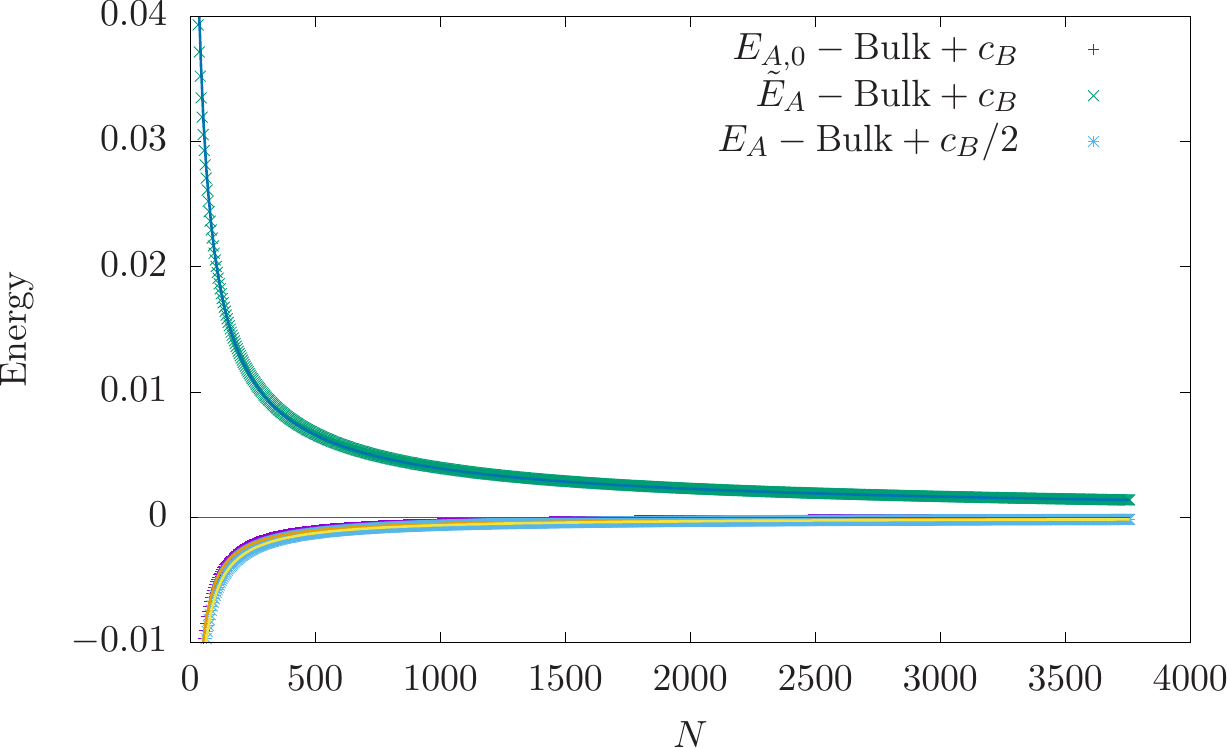}
  \includegraphics[width=8cm]{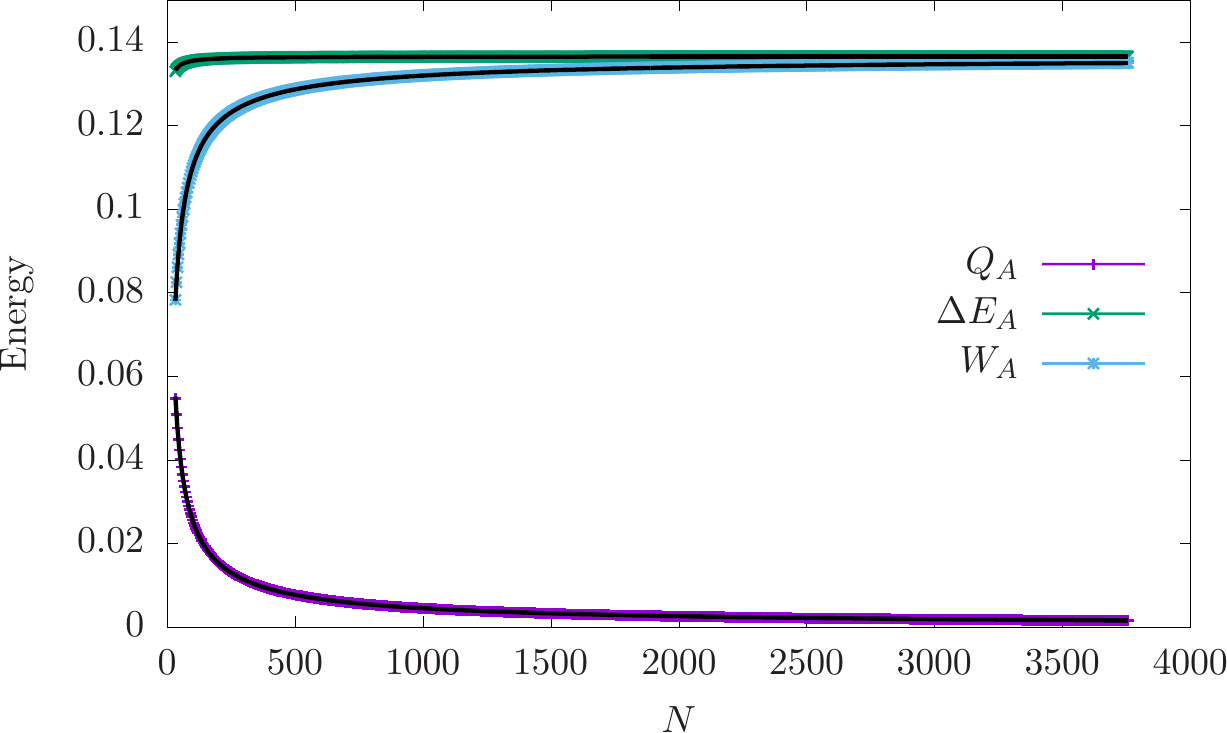}
  \caption{Top: The three energies involved, $E_A$, $\tE$ and $\Ez$,
    for a free fermionic chain, with A the left half, as a function of
    the system size, along with the theoretical asymptotic
    expressions, Eqs. \eqref{eq:EA0}, \eqref{eq:EA} and
    \eqref{eq:EAt}. Bottom: The three energy differences, $\Delta
    E_A=E_A-\Ez$, $W_A=E_A-\tE$, $Q_A=\tE-\Ez$, and their expected
    theoretical values according to Eqs. \eqref{eq:DEA}, \eqref{eq:Wt}
    and \eqref{eq:Qt}.}
  \label{fig:E}
\end{figure}

We can use a similar strategy to estimate $E_A$, but we should proceed
with care. Indeed, we can obtain $E_A$ numerically from the GS of the
whole chain, substracting the energy associated to the central link
and dividing by two,

\beq
E_A = {E_0(N)\over 2} - C_{N/2,N/2+1}.
\eeq
The first term can be easily estimated from \eqref{eq:EA0},

\beq
{E_0(N)\over 2} \approx -c_0 \({N\over 2}-1\) +{c_0\over 2}-{c_B\over 2} -
{\pi\over 24 N},
\eeq
and the second one can be found making use of \eqref{eq:corr}, giving
rise to an alternating behavior,

\beq
C_{n,n+1}\approx -{c_0\over 2}-{\pi\over 24(N+1)^2} +
{(-1)^n \over 2(N+1)\sin\({\pi(n+1/2) \over N+1}\)},
\eeq
which, since $N/2$ is even, reduces for the central link to 

\beq
C_{N/2,N/2+1} \approx -{c_0\over 2} -{\pi \over 24(N+1)^2} + {1\over 2(N+1)},
\eeq
yielding

\beq
E_A \approx -c_0\({N\over 2}-1\) - {c_B\over 2} - \({\pi\over
  24}+{1\over 2}\) {1\over N}.
\label{eq:EA}
\eeq
We notice that the bulk term is exactly the same as for $\Ez$, and the
boundary term is exactly half, as we would expect intuitively, since
this subsystem only possesses one bondary instead of two. We should
stress that a naive calculation would yield a Casimir correction
$\pi/(24N)$, but we obtain an additional contribution from the energy
associated to the central link. The validity of the approximations to
these two energies, $E_{A,0}$ and $E_A$, can be checked in Fig.
\ref{fig:E}.

Therefore, the excess energy, $\Delta E_A=E_A-\Ez$, is given by

\beq
\Delta E_A \approx {c_B\over 2} + \({\pi\over 8} - {1\over 2}\)
       {1\over N}.
\label{eq:DEA}
\eeq

\subsection{Bound energy and entanglement}

Extracting the maximal amount of work through unitary operators
reversibly is equivalent to minimizing the block energy while
preserving the full spectrum of the reduced density matrix. Thus, we
proceed to {\em align} the occupation eigenvectors with the
eigenstates of $H_A$, whose eigenvalues will be denoted by
$\{\ve^A_k\}$. The passive energy $\widetilde E_A$ can be written as

\beq
\widetilde E_A= \sum_{k=1}^\ell \nu_k \ve^A_k.
\label{eq:ff_passive}
\eeq

Since $E_A\leq \tE_A \leq \Ez$, it is reasonable to consider that the
passive energy $\tE_A$ will also present the same bulk term as in
Eq. \eqref{eq:EA0}, but with different corrections. Let us provide a
similar asymptotic expansion to its value.

\begin{figure}
  \includegraphics[width=8cm]{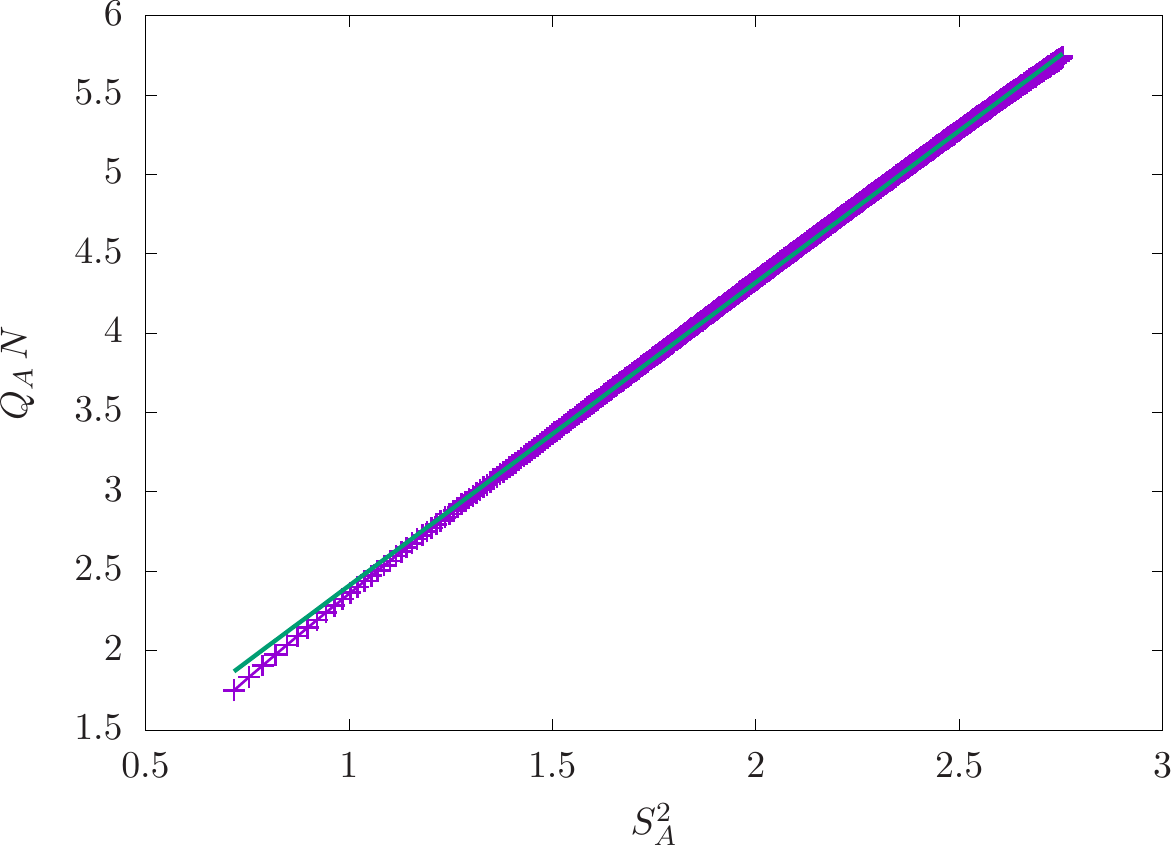}
  \caption{Numerical check of the linear relation between the bound
    energy multiplied by the system size, $Q_A\, N$, and the
    entanglement entropy squared, $S_A^2$ for the free fermionic
    chains, Eq. \eqref{eq:SQ}, for sizes $N$ in the same range as in
    Fig. \ref{fig:E}. The slope of the straight line, as expected, is
    $6/\pi\approx 1.9$.}
  \label{fig:SQ}.
\end{figure}

The eigenvalues of $H_A$ can be found exactly,

\beq
\ve^A_p = -2\cos \({p\pi \over N/2+1}\),
\eeq
with $p\in \{1\cdots N/2\}$, and those of the correlation matrix $C_A$
can also be approximated as

\beq
\nu^A_p \approx {1\over 1+\exp\(-\beta\(p-N/4\)\)},
\eeq
where $\beta$ corresponds to the entanglement gap, given in
Eq. \eqref{eq:Sgap} \cite{Tonni.18}. Thus, the passive energy is
given by

\beq
\tE_A = \sum_{p=1}^{N/2} \ve^A_p \nu^A_p \approx
\sum_{p=1}^{N/2}  {-2\cos (2\pi p/N) \over 1+e^{-\beta(p-N/4)}}.
\eeq
If we take the continuum limit, making use of the Sommerfeld expansion
\cite{Ashcroft} and the Euler-Maclaurin formula, we arrive at

\beq
\tE_A \approx -c_0\({N\over 2}-1\) - c_B - {c\pi v_F\over 12 N} +
   {2\pi^3 \over 3N \beta^2},
\eeq
so we obtain the final form

\beq
\tE_A \approx -c_0 \({N\over 2}-1\) - c_B - {\pi \over 6 N} +
   {\log^2(\gamma N) \over 6\pi N}.
   \label{eq:EAt}
\eeq
We may now find the analytic expression for the ergotropy,

\beq
W_A=E_A-\tE_A \approx {c_B\over 2} + \({\pi\over 8} -{1\over
  2}\){1\over N} -
{\log^2(\gamma N) \over 6\pi N},
\label{eq:Wt}
\eeq
where the requirement $W_A\geq 0$ demands that $c_B>0$. This expression
can be checked in the bottom panel of Fig. \ref{fig:E}. Furthermore,
we can estimate the bound energy,

\beq
Q_A=\tE_A - \Ez \approx {\log^2(\gamma N) \over 6\pi N}\geq 0,
\label{eq:Qt}
\eeq
which is unconditionally positive, and can also be checked in the
bottom panel of Fig. \ref{fig:E}. Notice that Eq. \eqref{eq:Qt}
implies that the bound energy is directly related to the inverse
squared of the entanglement gap of the system, or the square of the
entanglement entropy. Using Eq. \eqref{eq:Scft} and
Eq. \eqref{eq:Sgap}, we obtain an approximate relation

\beq
Q_A\, N \approx {6\over\pi} S_A^2,
\label{eq:SQ}
\eeq
which provides a relation between the entanglement entropy of a block
of a free fermionic chain and the bound energy
associated. Eq. \eqref{eq:SQ} is the main prediction of this work, and
we conjecture that its validity extends beyond the case of free
fermionic chains, to any critical state in 1D described by a conformal
field theory. The validity of this expression can be numerically
checked in Fig. \ref{fig:SQ}.

We may define an ergotropy fraction $w_A=W_A/\Delta E_A$ and a bound
fraction, $q_A=Q_A/\Delta E_A$, as the ratios between the ergotropy or
the bound energy to the excess energy. We can see that $w_A\to 1$ and
$q_A\to 0$ as $N\to\infty$, implying that for larger systems we can
extract most of the excess energy in the form of work using unitary
transformations.


\section{Preliminary results on other critical models}
\label{sec:interacting}

We have considered two other spin chains, the critical Ising model in
a transverse field (ITF) and the Heisenberg model, and performed
numerical explorations using a combination of Lanczos and exact
diagonalization for small systems which provide preliminary numerical
evidence of the validity of Eqs. \eqref{eq:Qt} and \eqref{eq:SQ} for
these systems.

\begin{figure}
  \includegraphics[width=8cm]{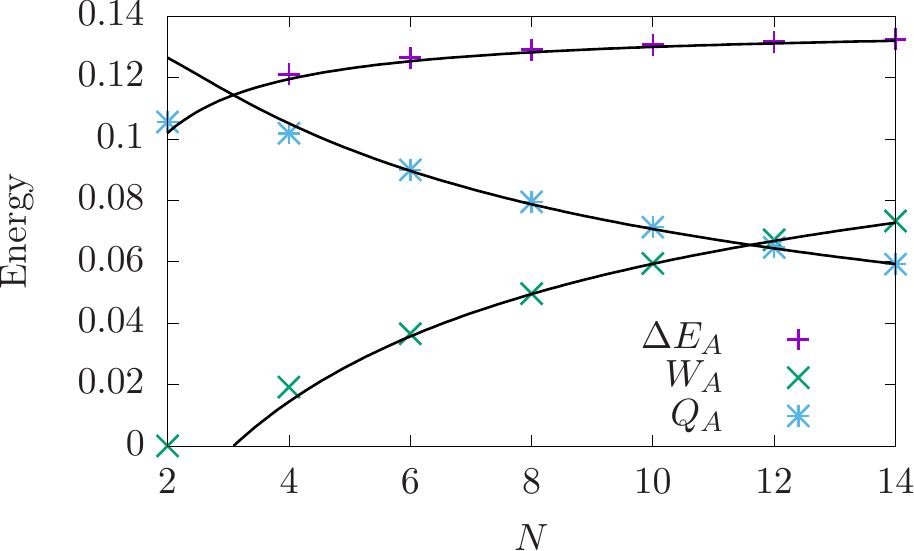}
  \includegraphics[width=8cm]{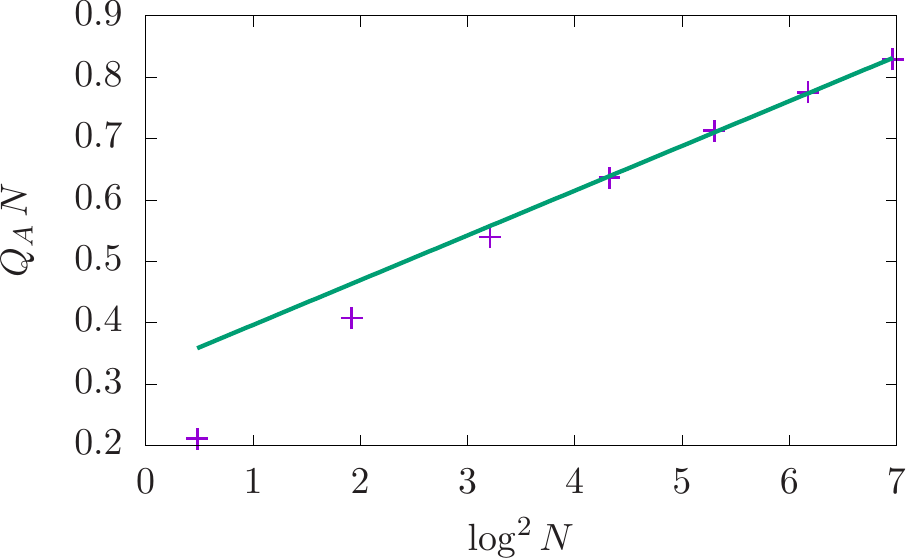}
  \caption{Subsystem energy decomposition for small Ising critical
    chains, with $N$ up to $14$. Top: Energies $\Delta E_A$, $W_A$ and
    $Q_A$ for the left-half chain as a function of the system size,
    along with the expected theoretical fits. Bottom: Approximate linear
    relation between $Q_A\, N$ and $\log^2(N)$, showing the
    expected relation between $Q_A$ and $S_A$, Eq. \eqref{eq:SQ},
    along with a linear fit to the last five points.}
  \label{fig:ising}
\end{figure}
    
\begin{figure}
  \includegraphics[width=8cm]{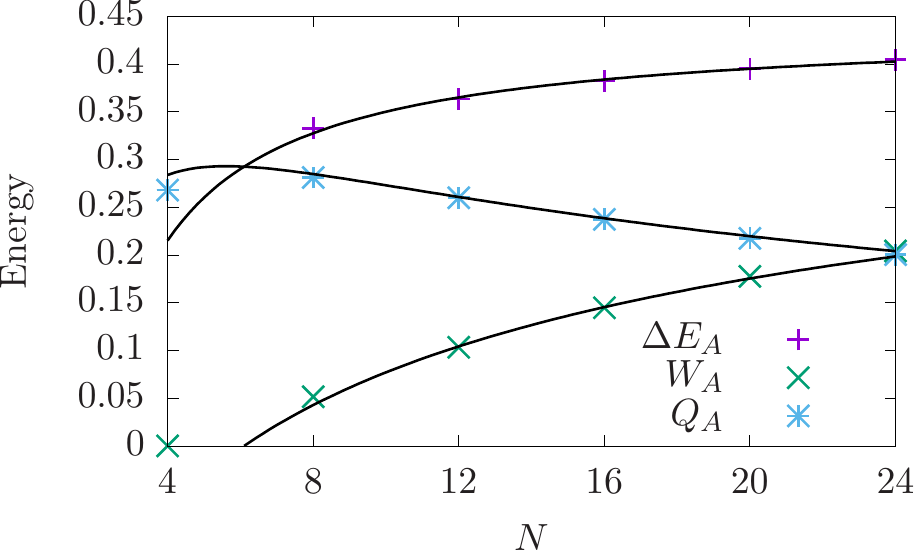}
  \includegraphics[width=8cm]{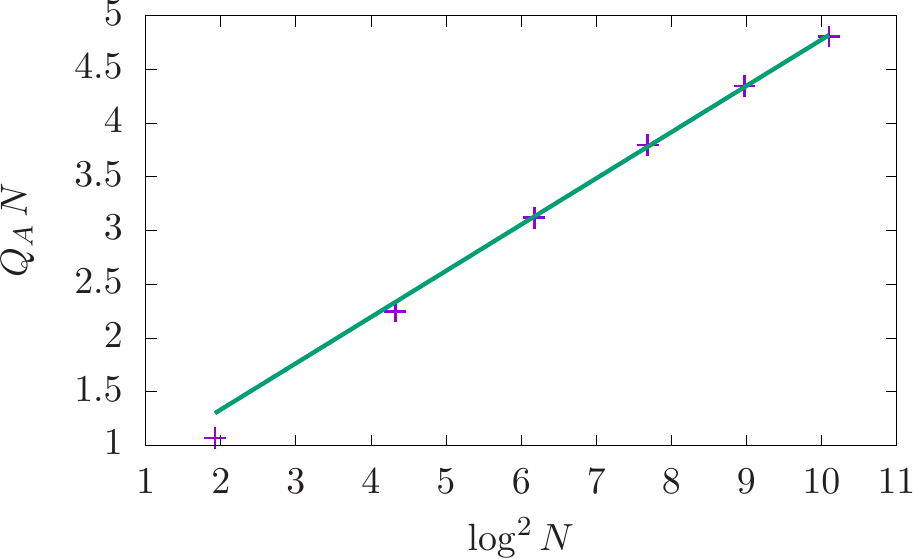}
  \caption{Subsystem energy decomposition for small Heisenberg chains,
    with $N$ up to 24, using only multiples of four. Top: Energies
    $\Delta E_A$, $W_A$ and $Q_A$ for the left-half chain, along with
    the expected theoretical fits; Bottom: Approximate linear relation
    between $Q_A\, N$ and $\log^2(N)$, showing the expected
    relation between $Q_A$ and $S_A$, Eq. \eqref{eq:SQ}, along with a
    linear fit to the last five points.}
  \label{fig:heis}
\end{figure}

The Hamiltonian of the ITF model that we have considered is given by

\beq
H_{\text{ITF}}=-\sum_{i=1}^{N-1} \sigma^z_i \sigma^z_{i+1} - \Gamma
\sum_{i=1}^N \sigma^x_i,
\eeq
for $\Gamma=1$. The low energy eigenstates of $H_{\text{ITF}}$ are
known to follow a conformal field theory with central charge $c=1/2$
\cite{DiFrancesco,Mussardo}. Therefore, the entanglement entropy of
the left half can be written as a linear function of $\log(N)$. We
have obtained preliminary numerical results employing exact
diagonalization up to size $N=14$, which are shown in Fig.
\ref{fig:ising}. In the top panel we show with points the energy
decomposition, $\Delta E_A$, $W_A$ and $Q_A$, for the left-half chain
of the even sized systems, along with their fits with continuous lines
to theoretical curves suggested by the generalization of Eqs.
\eqref{eq:DEA}, \eqref{eq:Wt} and \eqref{eq:Qt}, i.e.

\begin{align}
  \Delta E_A & \approx \alpha_1 -{\alpha_2\over N},\nn\\
  W_A &\approx \alpha_1 - {\alpha_2 \over N} - \alpha_3{\log^2(\alpha_4N)
    \over N},\nn\\
  Q_A &\approx \alpha_3 {\log^2(\alpha_4 N)\over N}.
  \label{eq:functional_form}
\end{align}
In our case the optimal values of the parameters are $\alpha_1\approx
0.137$, $\alpha_2\approx 0.07$, $\alpha_3\approx 0.044$ and
$\alpha_4\approx 5.5$. We would like to stress that we fit the 21
points of the three curves using the same values for the $\alpha_i$
parameters. In the bottom panel of Fig. \ref{fig:ising} we observe an
approximate linear relation between $NQ_A$ and $\log^2(N)$, as
expected, along with a linear fit obtained from the larger systems.
Even though the functional form is shown to be approximately correct,
we should use these fitting parameters with care, due to the small
system size.

\bigskip

On the other hand, we have considered the antiferromagnetic spin 1/2
Heisenberg chain with open boundaries, given by the Hamiltonian

\beq
H_{\text{Heisenberg}}=\sum_{i=1}^{N-1} \vec S_i \cdot \vec S_{i+1},
\eeq
which also corresponds to a conformal field theory for low energies,
with $c=1$ in this case, and can be mapped to an interacting fermion
Hamiltonian using the Jordan-Wigner transformation
\cite{DiFrancesco,Mussardo}. As it was mentioned above, the GS can be
analytically obtained using the Bethe Ansatz, but we have chosen to
obtain it using the Lanczos algorithm up to $N=24$, taking into
account the full SU(2) symmetry of the model. The top panel of Fig.
\ref{fig:heis} shows the energy decomposition for the left-half of the
chain, using only values of $N$ which are multiples of four. Again, we
plot along a fit of these 18 points to the form
\eqref{eq:functional_form}, obtaining approximate parameters
$\alpha_1\approx 0.44$, $\alpha_2\approx 0.9$, $\alpha_3\approx 0.41$
and $\alpha_4 \approx 1.32$. The bottom panel of Fig. \ref{fig:heis}
shows the linear relation between $Q_A\, N$ and $\log^2(N)$,
highlighting the validity of Eq. \eqref{eq:SQ}, again comparing to a
linear fit for the largest sizes.

The approximate validity of Eq. \eqref{eq:SQ} in all three models is
related to the fact that it only depends on the following facts:

\begin{itemize}
\item The Casimir expression for the energy of the GS.
\item The affine relation between the entanglement entropy and
  $\log(N)$.
\item The approximate inverse relation between the entanglement
  entropy and the entanglement gap.
\end{itemize}

All these relations stem from conformal invariance, a property
shared by all three models discussed in this work.

It would be interesting to check the validity of our preliminary
results for larger system sizes in the ITF and Heisenberg cases. The
ITF case can be evaluated using a combination of Jordan-Wigner and
Bogoliubov transformations. The Heisenberg case is more involved,
since e.g. the density matrix renormalization group (DMRG) can not be
used in a straightforward manner \cite{DMRG}, because we need to use
both the entanglement spectrum and the full energy spectrum of the
subsystem.


\section{Conclusions and further work}
\label{sec:conclusions}

In this work we have considered the excess energy possessed by a
subsystem of a ground state. Part of this excess energy can be
extracted via unitary operations, which we call {\em subsystem
  ergotropy}, and part of it can not be extracted in this way, which
we call {\em subsystem bound energy}. For concreteness, we have
considered one-dimensional systems which present conformal invariance,
and we have done the calculations in detail for free fermionic chains,
combining numerical calculations with a detailed analysis of the
Casimir corrections to the GS energy. The most relevant relation found
is a linear functional dependence between the subsystem bound energy
and the square of its entanglement entropy divided by the system size.
We have shown that this relation is likely to apply to other critical
spin chains, thus allowing us to conjecture that its validity will
extend to all 1D conformal field theories.

We would like to stress that, as the system size grows, the fraction
of excess energy which can be extracted as work approaches one. In
other words: almost all the subsystem energy becomes available in the
thermodynamic limit. This result is non-trivial, although it responds
to our intuition that for larger systems we have a larger freedom to
manipulate the local mixed state. It is relevant to ask how general
this result is. For instance, we may wonder about the behavior of the
subsystem ergotropy away from criticality, i.e. for dimerized spin
chains or for the Ising model with a non-critical value of the
transverse field $\Gamma$, or how to extend it to higher dimensional
systems.

Our results encourage further exploration of the application of
quantum thermodynamics to the analysis and characterization of
entanglement. Beyond the quantitative study of the ergotropy and bound
energies, it is relevant to ask about the passive state which we
obtain when all the ergotropy has been obtained. Indeed, it must be a
thermal state under the entanglement Hamiltonian, but it is also
relevant to ask about its properties under its own local Hamiltonian,
and how do these two Hamiltonians relate. Given the relation between
the entanglement Hamiltonian and the Unruh effect
\cite{Unruh.76,Takagi.86,Laguna.17}, this research programme may bear
fruits also to the interplay between gravity, entanglement and
thermodynamics.

\bigskip


\begin{acknowledgments}
  We acknowledge very useful discussions with G. Sierra, N. Samos
  Sáenz de Buruaga, K. Zawadzki, S. Singha Roy, J. Hoyos, G.M.
  Andolina and the anonymous referees. This work was funded by the
  Spanish government through grants PGC2018-094763-B-I00,
  PID2019-105182GB-I00 and PID2021-123969NB-I00 and by Comunidad de
  Madrid (Spain) under the Multiannual Agreement with UC3M in the line
  of Excellence of University Professors (EPUC3M14 and EPUC3M23), in
  the context of the V Plan Regional de Investigación Científica e
  Innovación Tecnológica (PRICIT). B.M. acknowledges financial support
  through contract No. 2022/167 under the EPUC3M23 line.
\end{acknowledgments}


\begin{thebibliography}{999}

  
\bibitem{Gemmer.04} J. Gemmer, M. Michel, G. Mahler, {\em Quantum
  Thermodynamics}, Springer (2004).

\bibitem{Binder.18} F. Binder, L.A. Correa, C. Gogolin, J. Anders,
  G. Adesso, {\em Thermodynamics in the quantum regime}, Springer
  (2018).
  
\bibitem{Bera.19} M.N. Bera, A. Riera, M. Lewenstein, Z.B. Khanian,
  A. Winter, {\em Thermodynamics as a consequence of information
    conservation}, Quantum {\bf 3}, 121 (2019).


\bibitem{Scully.03} M. Scully, M.S. Zubairy, G.S. Agarwal, H. Walther,
  {\em Extracting work from a single heat bath via vanishing quantum
    coherence}, Science {\bf 299}, 862 (2003).

\bibitem{Ghosh.19} A. Ghosh, V. Mukherjee, W. Niedenzu, G. Kurizki,
  {\em Are quantum thermodynamic machines better than their classical
    counterparts?}  Eur. Phys. J. Special Topics {\bf 227}, 2043
  (2019).

\bibitem{Fernandez.22} J.J. Fernández, {\em Optimization of energy
  production in two-qubit heat engines using the ecological function},
  Quantum Sci. Technol. \textbf{7}, 035002 (2022).



\bibitem{Allahverdyan.04} A.E. Allahverdyan, R. Balian,
  Th.M. Nieuwenhuizen, {\em Maximal work extraction from finite
    quantum systems}, Europhys. Lett. {\bf 67}, 565 (2004).

\bibitem{Alicki.13} R. Alicki, M. Fannes, {\em Entanglement boost for
  extractable work from ensembles of quantum batteries}, Phys. Rev. E
  {\bf 87}, 042123 (2013).

\bibitem{Campaioli.17} F. Campaioli, F.A. Pollock, F.C. Binder,
  L. Céleri, J. Goold, S. Vinjanampathy, K. Modi, {\em Enhancing the
    charging power of quantum batteries,} Phys. Rev. Lett. {\bf 118},
  150601 (2017).

\bibitem{Andolina.19} G.M. Andolina, M. Keck, A. Mari, M. Campisi,
  V. Giovannetti, M. Polini, {\em Extractable work, the role of
    correlations, and asymptotic freedom of quantum batteries},
  Phys. Rev. Lett. {\bf 122}, 047702 (2019).

\bibitem{Rosa.20} D. Rosa, D. Rossini, G.M. Andolina, M. Polini,
  M. Carrega, {\em Ultra-stable charging of fast-scrambling SYK
    quantum batteries}, JHEP {\bf 11}, 067 (2020).
  
\bibitem{Francica.17} G. Francica, J. Goold, F. Plastina,
  M. Paternostro, {\em Daemonic ergotropy: enhanced work extraction
    from quantum correlations}, npj Quantum Information {\bf 3}, 12
  (2017).

\bibitem{Francica.20} G. Francica, F.C. Binder, G. Guarnieri,
  M.T. Mitchison, J. Goold, F. Plastina, {\em Quantum coherence and
    ergotropy}, Phys. Rev. Lett. {\bf 125}, 180603 (2020).

\bibitem{Tirone.21} S. Tirone, R. Salvia, V. Giovannetti, {\em Quantum
  energy lines and the optimal output ergotropy problem},
  Phys. Rev. Lett. {\bf 127}, 210601 (2021).
  
\bibitem{Touil.22} A. Touil, B. Çakmak, S. Deffner, {\em Ergotropy
  from quantum and classical correlations}, J. Phys. A:
  Math. Theor. {\bf 55}, 025301 (2022).
  
\bibitem{Francica.22} G. Francica, {\em Quantum correlations and
  ergotropy}, Phys. Rev. E {\bf 105}, L052101 (2022).


\bibitem{Manzano.18} G. Manzano, F. Plastina, R. Zambrini, {\em
  Optimal work extraction and thermodynamics of quantum measurements
  and correlations}, Phys. Rev. Lett. {\bf 121}, 120602 (2018).
  
\bibitem{Solfanelli.19} A. Solfanelli, L. Buffoni, A. Cuccoli,
  M. Campisi, {\em Maximal energy extraction via quantum measurement},
  J. Stat. Mech. 094003 (2019).



  

\bibitem{Bisognano.75} J.J. Bisognano, E.H. Wichmann, {\em On the
  duality condition for a Hermitian scalar field,} J. Math. Phys. {\bf
  16}, 985 (1975).

\bibitem{Bisognano.76} J.J. Bisognano, E.H. Wichmann, {\em On the
  duality condition for quantum fields,} J. Math. Phys. {\bf 17}, 303
  (1976).


\bibitem{Li.08} H. Li, F.D.M. Haldane, {\em Entanglement spectrum as
  a generalization of entanglement entropy: identification of
  topological order in non-abelian fractional quantum Hall effect
  states}, Phys. Rev. Lett., {\bf 101}, 010504, (2008).
  

\bibitem{Peschel.99} I. Peschel, M.-C. Chung, {\em Density Matrices
  for a Chain of Oscillators,} J. Phys. A {\bf 32} 8419 (1999).

\bibitem{Chung.00} M.-C. Chung, I. Peschel, {\em On Density-Matrix
  Spectra for Two-Dimensional Quantum Systems}, Phys. Rev. B {\bf 62},
  4191 (2000).

\bibitem{Chung.01} M.-C. Chung, I. Peschel, {\em Density-Matrix
  Spectra of Solvable Fermionic Systems,} Phys. Rev. B {\bf 64},
  064412 (2001).

\bibitem{Peschel.03} I. Peschel, {\em Calculation of reduced density
  matrices from correlation functions,} J. Phys. A {\bf 36}, L205
  (2003);

\bibitem{Casini.09} H. Casini, M. Huerta, {\em Reduced density
  matrix and internal dynamics for multicomponent regions,}
  Class. Quant. Grav. {\bf 26}, 185005 (2009).

\bibitem{Casini.11} H. Casini, M. Huerta, R. Myers, {\em Towards a
  derivation of holographic entanglement entropy,} JHEP {\bf 1105},
  036 (2011).

\bibitem{Peschel.17} V. Eisler, I. Peschel, {\em Analytical results
  for the entanglement Hamiltonian of a free-fermion chain,}
  J. Phys. A: Math. Theor. {\bf 50}, 284003 (2017).

\bibitem{Santos.18} H. Santos, J.E. Alvarellos, J. Rodríguez-Laguna,
  {\em Entanglement detachment in fermionic systems}, Eur. Phys. J. D
  {\bf 72}, 203 (2018).

\bibitem{Samos.20} N. Samos Sáenz de Buruaga, S.N. Santalla,
  J. Rodríguez-Laguna, G. Sierra, {\em Piercing the rainbow:
    entanglement on an inhomogeneous spin chain with a defect},
  Phys. Rev. B {\bf 101}, 205121 (2020).
  
\bibitem{Cardy.16} J. Cardy, E. Tonni, {\em Entanglement hamiltonians
  in two-dimensional conformal field theory,} J. Stat. Mech. 123103
  (2016).
  
\bibitem{Tonni.18} E. Tonni, J. Rodríguez-Laguna, G. Sierra, {\em
  Entanglement Hamiltonian and entanglement contour in inhomogeneous
  1D systems}, J. Stat. Mech. 043105 (2018).
  
\bibitem{Samos.21} N. Samos Sáenz de Buruaga, S.N. Santalla,
  J. Rodríguez-Laguna, G. Sierra, {\em Entanglement in non-critical
    inhomogeneous quantum chains}, Phys. Rev. B {\bf 104}, 195147
  (2021).

  

\bibitem{DiFrancesco} P. di Francesco, P. Matthieu, D. Sénéchal, {\em
  Conformal Field Theory}, Springer (1997).

\bibitem{Mussardo} G. Mussardo, {\em Statistical Field Theory}, Oxford
  Graduate Texts (2010).

\bibitem{Holzhey.94} C. Holzhey, F. Larsen, F. Wilczek, {\em Geometric
  and renormalized entropy in conformal field theory}, Nucl. Phys.  B
  {\bf 424}, 443 (1994).

\bibitem{Vidal.03} G. Vidal, J. I. Latorre, E. Rico, A. Kitaev, {\em
  Entanglement in quantum critical phenomena}, Phys. Rev. Lett. {\bf
  90}, 227902 (2003).

\bibitem{Calabrese.09} P. Calabrese, J.Cardy, {\em Entanglement
  entropy and conformal field theory}, J. Phys. A: Math. Theor. {\bf
    42}, 504005 (2009).

  
\bibitem{Cardy.84} J.L. Cardy, {\em Conformal invariance and
  universality in finite-size scaling}, J. Phys. A: Math. Gen. {\bf
  17}, L385 (1984).

\bibitem{Cardy.86} H.W.J. Blöte, J.L. Cardy, M.P. Nightingale, {\em Conformal
  Invariance, the Central Charge, and the Universal Finite-Size
  Amplitudes at Criticality}, Phys. Rev. Lett. {\bf 56}, 742 (1986).

\bibitem{Mula.21} B. Mula, S.N. Santalla, J. Rodríguez-Laguna, {\em
  Casimir forces on deformed fermionic chains}, Phys. Rev. Research
  {\bf 3}, 013062 (2021).

\bibitem{Singha.20} S. Singha Roy, S.N. Santalla, J. Rodríguez-Laguna,
  G. Sierra, {\em Entanglement as geometry and flow}, Phys. Rev. B
  {\bf 101}, 195134 (2020).

\bibitem{Singha.21} S. Singha Roy, S.N. Santalla, J. Rodríguez-Laguna,
  G. Sierra, {\em Link representation of the entanglement entropies
    for all bipartitions}, J. Phys. A: Math. Theor. {\bf 54}, 305301
  (2021).

\bibitem{Zamora.14} A. Zamora, J. Rodríguez-Laguna,
  M. Lewenstein. L. Tagliacozzo, {\em Splitting a critical spin
    chain}, J. Stat. Mech. P09035 (2014).

\bibitem{Shimaji.19} T. Shimaji, T. Takayanagi, Z. Wei, {\em
  Holographic quantum circuits from splitting/joining local quenches},
  JHEP 165 (2019).

\bibitem{Rolph.22} A. Rolph, {\em Local measures of entanglement in
  black holes and CFTs}, SciPost Phys. {\bf 12}, 079 (2022).
  
\bibitem{Ashcroft} N.W. Ashcroft, N.D. Mermin, {\em Solid state
  physics}, Harcourt College Publ. (1976).

\bibitem{DMRG} U. Schollwöck, {\em The density-matrix renormalization
  group in the age of matrix product states}, Ann. Phys. {\bf 326}, 96 (2011).
  

\bibitem{Unruh.76} W.G. Unruh, {\em Notes on black-hole evaporation},
  Phys. Rev. D {\bf 14}, 870 (1976).

\bibitem{Takagi.86} S. Takagi, {\em Vacuum noise and stress induced by
  uniform accelerator: Hawking-Unruh effect in Rindler manifold of
  arbitrary dimensions}, Prog. Theor. Phys. Supp. {\bf 88}, 1 (1986).

\bibitem{Laguna.17} J. Rodríguez-Laguna, L. Tarruell, M. Lewenstein,
  A. Celi, {\em Synthetic Unruh effect in cold atoms}, Phys. Rev. A
  {\bf 95}, 013627 (2017).
  


\end{thebibliography}
\end{document}